\documentclass[runningheads,citeauthoryear]{apinv}
\usepackage{epsfig,cite,graphics}

\usepackage[T2A]{fontenc}
\usepackage[cp1251]{inputenc}
\usepackage[bulgarian,english]{babel}

\def\e{$\pm$}
\def\kms{km$\,$s$^{-\!1}$} 
\def\vsi{$v\: \sin i$}

\newcommand{\ltsima} {$\; \buildrel < \over \sim \;$} 
\newcommand{\simlt} {\lower.5ex\hbox{\ltsima}} % < over MMM 
\newcommand{\gtsima} {$\; \buildrel > \over \sim \;$} 
\newcommand{\simgt} {\lower.5ex\hbox{\gtsima}} % > over MMM 

\begin{document}

\title{Tidal Interaction in High Mass X-ray Binaries and Symbiotic Stars}
\titlerunning{High mass X-ray binaries and symbiotic stars}
\author{Radoslav K. Zamanov}
\authorrunning{R. Zamanov}
\tocauthor{R. Zamanov}
\institute{Institute of Astronomy, Bulgarian Academy of Sciences
	\email{rkz@astro.bas.bg}
}
%% \papertype can be "research report", "review", "conference talk", "conference poster",
%% \ "lecture at scientific seminar", "summary of dissertation",  etc.
\papertype{invited talk}
\maketitle
\begin{abstract}
This paper summarizes our recent results on tidal interaction in 
high mass X-ray binaries and symbiotic stars. We demonstrate that the giant in 
symbiotic stars with orbital periods $ \le 1200 \; d$ are co-rotating (synchronized). 
The symbiotics MWC~560 and CD-43$^0$14304 probably have high orbital eccentricity.
The giants in symbiotic binaries rotate faster than the field giants, likely
their rotation is accelerated by the tidal force of the white dwarf.

The giant/supergiant High mass X-ray binaries with orbital periods $\le 40 \; d$ are synchronized.
However the Be/X-ray binaries are not synchronized. In the Be/X-ray binaries the circumstellar disks 
are denser and smaller than those in isolated Be stars, probably truncated by the orbiting neutron star.
\end{abstract}
\Bg
\bgtitle{Приливно взаимодействие в масивни рентгенови двойни и симбиотични звезди}{
Радослав К. Заманов}{Тази статия представя резултатите от нашите изследвания на 
приливното взаимодействие в масивни рентгенови двойни и симбиотични звезди.
Ние демонстрираме, че симбиотичните звезди, с орбитални периоди $ \le 1200$ дни, са синхронизирани. 
Симбиотичните звезди MWC~560 и CD-43$^0$14304 вероятно имат орбити с голям ексцентрицитет.
Гигантите в симбиотични се въртят по-бързо от гигантите от полето, вероятно тяхното въртене е ускорено от приливната сила 
на бялото джудже.  \\
Масивните ретгенови двойни с гигант/свръхгигант и орбитални периоди $\le 40$ дни са синхронизирани. 
Ве/рентгеновите двойни не са синхронизирани. В тях околозвездните дискове са по-плътни и по-малки от тези в изолираните звезди,
вероятно отрязани от неутронната звезда. 
}
\Eng
\section{Introduction}

Stars in close binary systems are subject to mutual tidal forces that distort their
stellar shape, breaking their spherical and axial symmetry, which leads to different
observational effects -- ellipsoidal
variability and apsidal motion, 
circularization, transition period between circular and eccentric orbits,
synchronization and spin-orbit alignment
(e.g. Mazeh 2008). We investigate a few observational appearances of
the tidal force of the compact object on the mass donating star
in  symbiotic stars (SSs) and High Mass X-ray Binaries (HMXrB).

The symbiotic stars (thought to comprise  a white dwarf                    
accreting from a  cool giant or Mira) represent the  
extremum of the interacting binary star classification 
(e.g. Corradi et al. 2003).               		    
On the basis of their IR properties, SSs have been classified into stellar continuum (S) 
and dusty (D or D') types (Allen 1982). The D--type systems contain Mira variables as mass donors.  
The D'--type  are  characterized by an earlier spectral type (F-K) of the cool component
and lower  dust temperatures. 

In HMXrB a neutron star or stellar mass black hole accretes material from a massive early type star.
They are divided in two groups: (1) Be/X-ray binaries and (2) giant/supergiant systems.

The aims of our investigations are: \\
--  to measure the projected rotational velocities (\vsi)  
     and  the rotational periods (P$_{rot}$) of the  giants in a number of southern symbiotic stars;  \\
-- to perform comparative analysis and  explore theoretical predictions 
     that the mass donors in symbiotics are faster rotators compared with  field giants; \\
-- to check whether the rotation  of the red giants in SSs 
     is synchronized with the orbital period; \\
-- to check whether the rotation of the mass donors in 
HMXrB is synchronized with the orbital period.

\section{Observations}
We have observed 42 symbiotic stars --
all southern  S- and D'-SSs  from the catalogue 
Belczy{\' n}ski et al. (2000) with $0^h<RA<24^h$, 
declination $<2^0$, and catalogue magnitude brighter than $V<12.5$.

The observations have been performed  with FEROS at the 2.2m 
telescope (ESO, La Silla).   
FEROS is a fibre-fed echelle spectrograph, providing a high resolution of 
$\lambda/\Delta \lambda=$48000, 
a wide wavelength coverage from about 4000~\AA\  to 8900~\AA\  in one exposure 
(Kaufer et al. 1999). All spectra are reduced using the dedicated FEROS data reduction software 
implemented in the ESO-MIDAS system. Using CCF and FWHM methods, we measured 
the projected rotational velocity (\vsi) on our observations,
and on spectra from the archives of VLT/UVES and ELODIE (Zamanov et al. 2008). 

\section{The mass donors in S--type symbiotics are co-rotating}  
\label{Sec_sta}

%%%------------------------------------------------------------------------------  
 \begin{figure*}[!htb]
 \vspace{8.8cm}  
  \includegraphics{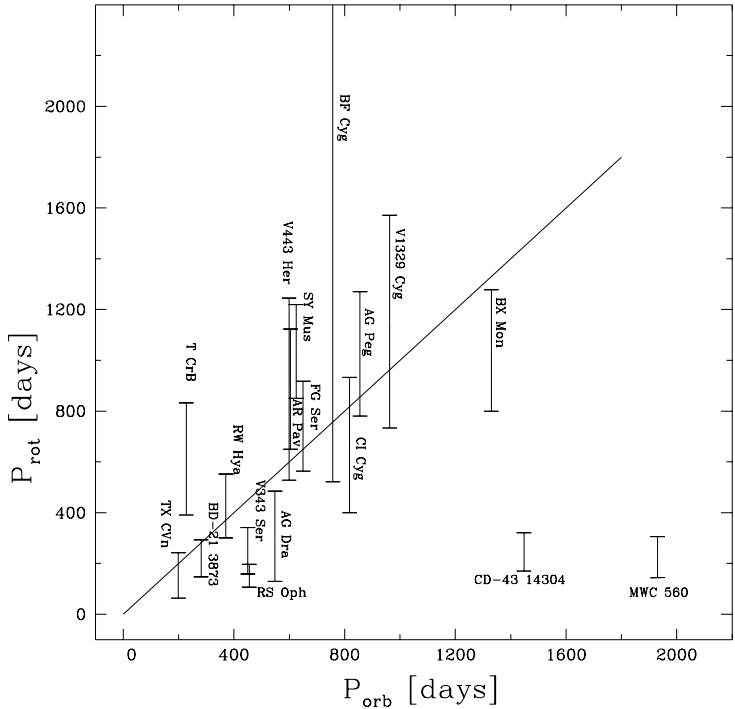}   
  \caption[]{The rotational period of the red giant (P$_{rot}$) 
  versus the orbital period (P$_{orb}$) for S-type symbiotics.  The solid line corresponds to 
  P$_{rot} =$P$_{orb}$.  Most objects are close to this line,
  which indicates that they are synchronized. There are 3 objects
  which deviate considerably from that rule (RS~Oph, CD-43$^\circ$14304, MWC~560).
  }		    
\label{P-P}     
\end{figure*}	      
%-------------------------------------------------------------------------------  

From our \vsi\ measurments and data collected from the literature, 
we calculate P$_{rot}$ (rotational period of the giant): 
\begin{equation}
 P_{rot}=\frac { 2 \pi R_g \sin i  }{v\: \sin i },
\end{equation}
where $R_g$ is the radius of the mass donor, $i$ is the inclination of the orbit to the line of sight.
In the most cases  for $R_g$, we use the the average radius for the corresponding spectral type taken from 
van Belle et al.(1999).

We collected 18 S-type SSs in total, for which we know the orbital period (P$_{orb}$) and P$_{rot}$.
Fig. \ref{P-P} shows P$_{rot}$ versus P$_{orb}$ of the 18 objects in our sample, 
with a straight line indicating the co-rotation (i.e. P$_{rot}$=P$_{orb}$). 
Most objects are close to this line, which suggests that they are synchronized. 
9 objects are synchronized within the measurement errors (1-$\sigma$ level).
4 objects have deviations between 1 and  2-$\sigma$. Generally, 15 out of 18 are  
within the 3-$\sigma$ level. 

The objects that deviate significantly from the P$_{orb}$=P$_{rot}$ line are
RS~Oph, MWC~560 (peculiar symbiotics), and CD-43$^\circ$14304. 

In Fig. \ref{P-P} figure, it is visible that the symbiotic stars with P$_{orb} < 1200$ days are synchronized (Zamanov et al. 2007).  
The statistical tests demonstrate that the deviations are (most probably) due to measurement errors
and not to intrinsic scatter. In other words the null hypothesis that all S-type SSs
with well measured \vsi\ are synchronized 
(excluding RS~Oph, CD-43$^\circ$14304, MWC~560) cannot be rejected at the 99\% confidence level.

\section{Orbit eccentricity of MWC~560 and CD-43$^0$14304}

MWC~560 (V694~Mon)  is a symbiotic binary system,
which consists of a red giant and a white dwarf   
(Michalitsianos et al. 1993).
The most spectacular features of this object are the collimated ejections 
of matter with velocities of up to $\sim 6000$~\kms\ (Tomov et al. 1992; Stute \& Sahai 2009)
and the resemblance of its emission line spectrum to that of 
the low-redshift quasars.
The jet ejections are along the line of sight and the 
system is seen almost pole-on ($i < 16^\circ$).

In a binary with a circular orbit the rotational period of the primary, P$_{\rm rot}$, 
reaches an equilibrium value at the orbital period, $P_{\rm rot} = P_{\rm orb}$.  
However, in a binary with an eccentric orbit, the 
tidal force acts to synchronize the rotation of the mass donor with 
the motion of the compact object at the periastron -- the effect called 
pseudosynchronous rotation (Hall 1986).
The corresponding equilibrium (i.e. pseudosynchronization) is reached 
at a value of $P_{\rm rot}$ which is
{\it less} than $P_{\rm orb}$, the amount less being a function of the orbital
eccentricity $e$. 
Hut (1981) showed that the period of pseudosynchronization, P$_{\rm ps}$, is :

\begin{equation}
P_{\rm ps} = \frac{(1+3e^2+\frac{3}{8}e^4)(1-e^2)^\frac{3}{2}}{1+\frac{15}{2}e^2+
\frac{45}{8}e^4+\frac{5}{16}e^6} P_{\rm orb}.
\label{Eq-ps}
\end{equation}
For MWC~560, we measured \vsi $= 8.2 \pm 1.0$~\kms.  Following Schmid et al. (2001),
we adopt  red giant radius $R_{\rm g} = 140 \pm 7~R_\odot$, and  inclination 
$\; i=12^\circ-16^\circ$. We calculate $P_{\rm rot}=155 - 270$~days. This value 
is considerably less than the orbital period, $P_{\rm orb}=1931 \pm 162$~days 
($P_{\rm orb}$ from Gromadzki et al. 2007).
Following the estimations of the timescales (Stoyanov 2010), MWC~560 should 
be close to synchronization or pseudosynchronization, and $P_{\rm rot} =P_{\rm ps}$. 
Using Eq.~\ref{Eq-ps} we can therefore estimate the orbital eccentricity to be 
$e=0.73-0.79$.

{\bf CD-43$^0$14304} : It is a symbiotic star with mass donor K5III 
and a white dwarf with temperature T=110~000~K.
We assume  $P_{\rm orb}= 1448 \pm 100$~days (Schmid et al., 1998) 
and red giant radius $R_{\rm g} = 38.8 \pm2 ~R_\odot$.  
Spectropolarimetry   (Harries \& Howarth 2000) gives two possible values
for the inclination of the system $i=57^0\pm5^0$  or $i=122^0\pm48^0$. 
Both values imply $\sin i \approx 0.84$. We calculate 
$P_{\rm rot} \approx 170 - 321$~d and $e$ (using Eq.~\ref{Eq-ps}).
It is likely that the eccentricity is high $e=0.61-0.76$.

\subsection{D'-type symbiotics}

There are  7 D' SSs listed in the catalogue of Belczy{\' n}ski et al. (2000).
Rotational velocities are  measured for all six southern objects. They are summarized in Table~\ref{tab_D'}.
\vsi\ of V471~Per is unknown, but it will be valuable to measure it.

\begin{table}  
\caption{D'-type symbiotics. In the table are gives as follows: name of the object,
spectral type of the mass donor, \vsi\ (our measurement), \vsi\ of field giants with identical spectral
type, adopted radius of the giant ($R_g$), critical velocity, the ratio $\frac{v\: \sin i}{v_{\rm crit}}$. }  
% \begin{tabular}{@{}llr| c |rr | rrr @{}}  
\begin{tabular}{lclr| c |rr | rrr}  
\hline
\hline
Object          & IR   &   Spec	& \vsi  	 &  similar objects & $R_g$	  & v$_{crit}$   &   $\frac{v\: \sin i}{v_{\rm crit}}$  \\
                & type &    	& [\kms]	 &   \vsi\ [\kms]   & [R$_\odot$] &  [\kms]	 &		    \\
&&& &&& \\
\hline
&&& &&& \\
WRAY~15-157     & D' &   G5III	  &    37\e5.0	   &  $< 15$     & 10.0 & 198 & 0.19	    \\
HD 330036       & D' &   F8III	  &    107.0\e10   &  10-35      & 22.1 & 160 & {\bf 0.67}  \\   
Hen 3-1591      & D' &   K1III	  &    23.7\e 2.0  &  1-41       & 23.9 & 144 & 0.16	    \\
StH$\alpha$ 190 & D' &   G4III/IV &    105.0\e10   &  $<24$      & 7.88 & 191 & {\bf 0.54}  \\
V417 Cen        & D' &   G9Ib-II  &    75.0\e 7.5  &  1-20       & 75.0 & 105 & {\bf 0.71}  \\  
AS~201          & D' &   F9III	  &    29.3\e3.0   &  10-35      & 24.5 & 150 & 0.19	    \\	
V471~Per        & D' &   G5III	  &      ?	   &  $< 15$     & 10.0 & 198 &	--	    \\
Hen 3-1674      & S  &   M5III	  &    52.0\e6     &  $<20$      & 139.6& 60  & 0.8 ?       \\
\hline 						 
 \label{tab_D'} 
 \end{tabular}									   				      
 \end{table}	

{\bf WRAY~15-157}: The  catalogue of rotational  velocities for evolved stars (de Medeiros \& Mayor  1999)  
lists 18  objects from  spectral type G5III.  
They all rotate with \vsi$<15$ \kms.
WRAY~15-157 with \vsi$=37\pm5$ \kms, is an extremely  fast rotator for this spectral class.

{\bf Hen 3-1591} 
The same catalogue lists $>$60 K1III stars, and  90\% of them rotate with \vsi$<$8 \kms.
There are only 5 with \vsi$>$20 \kms. This means that
Hen~3-1591 is a very fast rotator (in the top 5\%). 

{\bf AS 201, HD~330036}: The same catalog  contains 5 objects from 
spectral type F8III-F9III. They rotate with \vsi\ of 10-35 \kms. 
AS 201 is well within in this range. However HD~330036 is an extremely
fast rotator. The same catalog lists 60 objects from 
spectral type G3,G4,G5 III-IV. 
They all rotate with \vsi\ $<24$ \kms. 
Again, this means that StH$\alpha$~190 is an extremely fast rotator. 

{\bf V417~Cen}: The catalog of de Medeiros et al. (2002) of \vsi\ of Ib supergiant stars
contains 16 objects from spectral type G8-K0 Ib-II. All they have
\vsi\  in the range 1-20 \kms. It means that V417~Cen
is an extreme case of very fast rotation for this spectral class.

{\bf V471~Per}: \vsi\ of V471~Per is unknown, but it will be valuable to measure it.

{\bf Hen 3-1674} is classified as S-type, however its rotation is similar to that of D'-SSs. 
an independent check of the parameters will be valuable.

There is a natural upper
limit for rotation speeds, where the centripetal acceleration balances that due to
gravitational attraction, often named the ``critical
speed'', where $v_{\rm crit} = \sqrt{GM/1.5R} = 357\sqrt{M/R}$ \kms\
(the factor of 1.5 appears from the assumption that at critical rotation speeds the
equatorial radius is 1.5 times the polar radius, $R$). 
The calculated $v_{\rm crit}$ is included in Table~\ref{tab_D'}.
No star can rotate faster than its critical speed,
however we can see that at least three D'--type SSs are rotating  at a substantial fraction of
their critical speeds. 
For the remaining  objects we can not exclude the possibility that
they also rotate very fast but are observed at low 
inclination ($i \le 30^0$).

All mass donors in D'--type systems appeared to be very fast rotators (see also Pereira et al. 2005;
Zamanov et al. 2006).
Four  of them are the fastest rotators in their spectral class.

As an after effect of the fast rotation D'-SS should be flattened at the poles and bulging at the equator.
The mass loss rate must be enhanced in equatorial regions, 
this will create dense circumstellar disk. In the outer parts of the disk there should be 
conditions for formation of dust.
Consequently the appearance of the dust in these systems is  (probably) 
a direct consequence of their fast rotation. 

If these binary stars are synchronized, their orbital periods 
should be relatively short (4-60 days).

\section{Fast rotation in S-type symbiotics}

Soker (2002) has predicted theoretically that the cool companions in symbiotic systems  
are likely to rotate much faster than isolated cool giants or those in wide binary 
systems.  Our observational investigation (Zamanov et al. 2006, 2008) 
clearly confirms theoretical predictions 
that the mass donors in symbiotics are fast rotators.

%----------------------------------------------------
\begin{table}
\caption{Projected rotational velocities of K and M giants.
 In the table are given as follows: the spectral type,  
 the  mean projected rotational  velocity (\vsi, in \kms), 
 standard deviation of the mean ($\sigma$, in \kms), the number of objects.
 In the second column are given the values for the field M giants, 
 in the third - for the symbiotic stars. }
\begin{tabular}{rr| r r | rrr|rrr}
\hline
\\
Spectral     &   &   {\bf field giants}     &   & & {\bf symbiotics}  &   & \\
bin          &   &   mean \vsi \e $\sigma$  & N & &  mean\e $\sigma$  & N & \\ 
             &   &   [\kms]                 &   & & [\kms]            &   & \\
\hline
&&& &&&                    \\
K2-K5   III  &   &  1.1\e1.4  & (363)	       & & 9.5\e6.7  & (7)  & \\
M0-M1   III  &   &  3.7\e1.9  & (23)	       & & 9.9\e2.6  & (2)  & \\
M1.5-M2 III  &   &  4.8\e4.1  & (14)	       & & 8.3\e1.1  & (3)  & \\
M2.5-M3 III  &   &  5.5\e2.0  & (8)	       & & 6.5\e1.8  & (4)  & \\
M3.5-M4 III  &   &  2.2\e1.0  & (3)	       & & 7.7\e3.3  & (7)  & \\
M4.5-M5 III  &   &  5.5\e4.0  & (5)	       & & 7.9\e1.7  & (9)  & \\
M5.5-M6 III  &   & 12.1\e5.1  & (4)	       & & 7.6\e2.0  & (6)  & \\
&&& &&& \\
 \hline
 \label{t-mean-sig}                                                      
 \end{tabular}        
\\                                              
 \end{table}	  
%-----------------------------------------------------------
The mean values of \vsi\ for the K and M giants are presented in 
Table~\ref{t-mean-sig}. All but one of the mean \vsi\ values of SSs 
are higher than those of the field giants. 

The K giants in S-type symbiotics rotate at \vsi$\: >\!4.5$~\kms, which is
2-- 4 times faster than the field K giants.
The majority of the field M giants rotate at about \vsi\ $\sim1-6$ \kms,
while the symbiotic M giants rotate at  \vsi\ $\sim 4 - 14$ \kms. 
The M giants in S-type symbiotics rotate on average 1.5 times faster than 
the field M giants. 

A few histrograms comparing symbiotics and field giants  can be seen in Zamanov et al. (2008). 
Statistical tests (Kolmogorov-Smirmov and Mann-Whitney U-test) show that these differences are highly 
significant -- p-value $< 10^{-3}$  in the spectral type bins K2III-K5III, M0III-M6III, 
and M2III-M5III;

%%%------------------------------------------------------------------------------  
 \begin{figure} % [!htb]
 \vspace{7.5cm}  
  \includegraphics{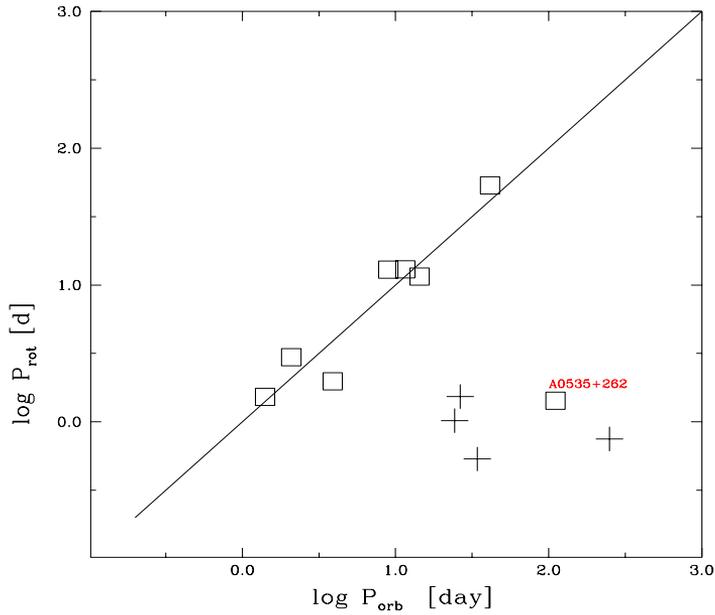}   
  \caption[]{The rotational period of the mass donor (P$_{rot}$) 
  versus the orbital period (P$_{orb}$) for 12 HMXrB.  The solid line corresponds to 
  P$_{rot} =$P$_{orb}$.  The squares are giant and supergiant systems, the crosses - Be/X-ray binaries.
  The giants/supergiants  are close to the line of synchronization.  
  The Be/X-ray binaries are not synchronized. 
  Typical errors are of the size of the symbols.
  }		    
\label{HMXrB}     
\end{figure}	      
%-------------------------------------------------------------------------------  

%%%------------------------------------------------------------------------------  
 \begin{figure*}[!htb]
 \vspace{8.4cm}  
  \includegraphics{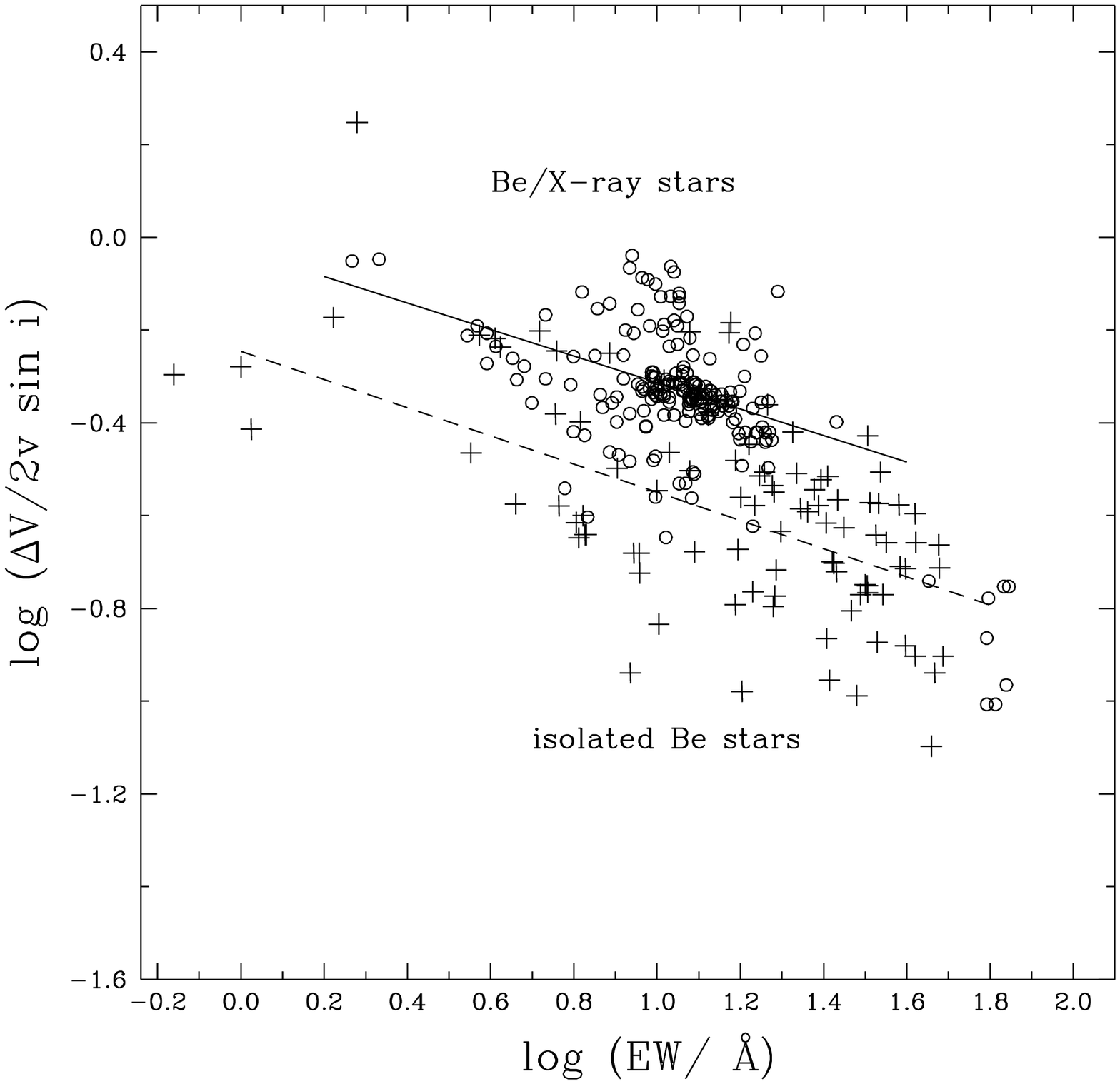}   
  \caption[]{Plot of H$\alpha $ line parameter $\log{(\Delta V/(2\:v\sin{i}))}$ 
 versus EW(H$\alpha $). The lines represent the best linear fits: the solid line 
 over the circles (Be stars), 
 the dashed line over the crosses (Be/X-ray stars). 
 The best fit line of the Be/X-ray binaries is shifted to denser circumstellar disks.
 Our estimation is that the circumstellar disks in the Be/X-ray systems 
 are about $\simeq$2 times more dense than disks in isolated Be stars. }		    
\label{BeXBe}     
\end{figure*}	      
%-------------------------------------------------------------------------------  

\section{High mass X-ray binaries}

\subsection{Synchronization}

We investigate the tidal interaction
in  High-Mass X-ray Binary stars in order to determine in 
which objects the rotation of the mass donors 
is synchronized or pseudosynchronized with 
the orbital motion of the compact companion. 
We calculate the rotation (P$_{rot}$) of the mass donor  and compare 
it with the orbital period (P$_{orb}$). The results for 12 HMXrB
with known orbital and stellar parameters are plotted in Fig.\ref{HMXrB} (see also Stoyanov \& Zamanov 2009)
 
We find that: {\bf (1)} the Be/X-ray binaries are not synchronized, 
the mass donors rotate faster than the orbital period;
{\bf (2)} the giant and supergiant systems are close to synchronization (at least 
for systems with orbital periods P$_{orb} < 40$ days). The only exception  
is 1A 0535+262 (V725 Tau). This object in its observational behaviour is more similar 
to the Be/X-ray binaries (e.g. Coe et al. 2006), and our result gives clue 
that the physical reason is the rotation of the mass donor.

\subsection{Comparison of the circumstellar disks in Be/X-ray binaries and Be stars}

We performed  a comparative study of the circumstellar disks in Be/X-ray binaries 
and isolated Be stars based upon the H$\alpha$ emission line (Zamanov et al. 2001). 
From this comparison it follows that the overall structure of the disks 
in the Be/X-ray binaries is similar to the disks of other Be stars, i.e. they are axisymmetric 
and rotationally supported. 
The factors for the line broadening (rotation and temperature) in the disks 
of the Be stars and the Be/X-ray binaries seem to be identical. 

However, we do detect some intriguing differences between the envelopes. 
On average, the disks in Be/X-ray binaries  have on average a smaller size,  
probably truncated by the compact object.
The different distribution of the Be/X-ray binaries and the Be stars seen in 
the normalized peak separation  versus equivalent width of $H \alpha$  diagram 
(see Fig. \ref{BeXBe})
indicates that the circumstellar disks of the Be/X-ray binaries are twice 
as dense as the disks of the isolated Be stars.

\section{Conclusions}
Our main results are as follows:

The giants in S- and D'-type symbiotic stars are fast rotators in comparison with field giants. 
At least three of the D'-symbiotic stars rotate at a substantial fraction of
the critical velocity. If D'-symbiotics are tidally synchronized, their orbital periods 
could be surprisingly short (5-50 d).

In the symbiotic stars with P$_{orb} < 1200$~d, the rotation of the 
red giant is tidally locked with the orbital motion.
Assuming  pseudosynchronization, we calculate that 
the orbital eccentricities of MWC~560 and CD-43$^0$14304 can be surprisingly large
($e$\simgt$0.6$).

% - As a consequence of the fast rotation the symbiotic giants should have 
% higher mass loss rate than isolated giants. 
% - As a result of the fast rotation 

In HMXrB, the rotation of the giants/supergiants is tidally synchronized 
for systems with for systems with $P_{orb}$ shorter than  $42 \;d$.

Nor the rotation of the  Be stars in Be/X-ray binaries, nor the formation of the circumstellar disk  is  
influenced by the  neutron star. However the circumstellar disks are truncated, in other words the influence
of the neutron is detected in the outer part of the circumstellar disks.   \\
\\
{\bf Open questions}, which should be addressed in the future: \\
- what is the reason for the extremely fast  rotation of D'-symbiotics? \\
- what is the case of Hen 3-1674?  \\
- are the symbiotics with P$_{orb} > 1200$~d synchronized? \\
- are the supergiants in HMXrB with P$_{orb} > 40$~d synchronized? \\

%\newpage

\end{document}